\documentclass[10pt,journal,cspaper,compsoc]{IEEEtran}

\usepackage{color}
\usepackage{cite}
\usepackage{graphicx}
\usepackage{algorithm}
\usepackage{algorithmic}
\usepackage{listings}
\usepackage{multirow}
\usepackage{epsfig}
\usepackage{times}
\usepackage{mathptm}
\usepackage{latexsym}
\usepackage{pslatex}
\usepackage{caption}
\usepackage{subcaption}
\usepackage{balance}

\begin{document}

\title{An Algorithm for Constructing a Smallest Register with Non-Linear Update Generating a Given Binary Sequence}%\titlenote{}}

\author{
Nan~Li,
~\IEEEmembership{Student Member,~IEEE}
and
Elena~Dubrova,
~\IEEEmembership{Member,~IEEE}
\thanks{The authors are with the Royal Institute of Technology (KTH), Stockholm, Sweden.

}
}

\IEEEcompsoctitleabstractindextext{
\begin{abstract}

Registers with Non-Linear Update (RNLUs) are a generalization of Non-Linear Feedback Shift Registers (NLFSRs) in which both, feedback and feedforward, connections are allowed and no chain connection between the stages is required.
In this paper, a new algorithm for constructing RNLUs generating a given binary sequence is presented.
Expected size of RNLUs constructed by the presented algorithm is proved to be $O(n/\log_2(n/p))$,
where $n$ is the sequence length and $p$ is the degree of parallelization.
This is asymptotically smaller than the expected size of RNLUs constructed by previous algorithms and
the expected size of LFSRs and NLFSRs generating the same sequence.
%It is also shown that, if a smaller RNLU generating the same sequence as the RNLU constructed by the presented algorithm exists, then their expected sizes differ by no more than $O(log(n/p))$ gates.
%We also show our results are within $O(\log_2(n/p))$ gates from optimal.
The presented algorithm can potentially be useful for many applications, including testing, 
wireless communications, and cryptography.
\end{abstract}

}

\maketitle

\section{Introduction}
\label{sec:intro}

Binary sequences are important for many areas, including cryptography, wireless communications, and testing.   

In cryptography, pseudo-random binary sequences are used in stream cipher-based  {\em encryption}. A stream cipher produces a keystream by combining
a pseudo-random sequence with a message, usually by the bit-wise addition~\cite{robshaw94stream}. 
The security of stream ciphers 
is directly related to statistical properties of pseudo-random sequences. At present, there is no secure method for
generating  pseudo-random sequences which satisfy the extreme limitations of technologies like RFID. 
Low-cost RFID tags cannot dedicate more than a few hundreds of gates for security functionality~\cite{rfid_review}. Even the most compact of
today's encryption systems contain over 1000 gates~\cite{GoB08}. The lack of adequate protection mechanisms gives rise to many security problems and blocks off a variety of potential applications of RFID technology. 

In wireless communications, pseudo-random sequences are used for {\em scrambling} and {\em spreading} of the transmitted signal.
Scrambling is performed to give a transmitted signal some useful engineering properties, e.g. to reduce the probability of interference with adjacent channels or to simplify timing recovery at the receiver~\cite{LeK01}.  
Spreading increases a bandwidth of the original signal making possible to maintain, or even increase, communication performance when signal power is below the noise floor~\cite{PiSM82}. 
For both, scrambling and spreading, it is important to select pseudo-random sequences carefully, because their length, bit rate, correlation and other properties determine the capabilities of the resulting systems. Today's wireless communication systems typically use Linear Feedback Shift Register (LFSR) sequences, or sequences obtained by linearly combining pairs of LFSR sequences, such as Gold codes~\cite{Go67}.  
There are many theoretical results demonstrating the advantages of using nonlinear sequences in wireless communications. For example, complementary sequences  can solve the notorious problem of power control in Orthogonal Frequency Division Multiplexing (OFDM) systems by maintaining a tightly bounded peak-to-mean power ratio~\cite{DaJ98}. Popovich~\cite{popovich} has shown that multi-carrier spread spectrum systems using complementary and extended Legendre sequences outperform the best corresponding multi-carrier Code Division Multiple Access (CDMA) system using Gold codes. However, due to the lack of efficient hardware methods for generating nonlinear sequences, their theoretical advantages cannot be utilized at present.  

{\em Built-In-Self-Test} (BIST) uses the pseudo-random binary vectors usually generated on-chip by an LFSR as test patterns~\cite{McC85}. The hardware cost of an LFSR-based BIST is low. However, the test time of BIST may be long due to random-pattern resistant faults. Several methods for coping with these faults have been proposed, including modification of the circuit under test~\cite{EiL83}, insertion of control and observe points into the circuit~\cite{RaTKM04}, modification of the LFSR to generate a sequence with a different distribution of 0s and 1s~\cite{ChM84}, and generation of top-off test patterns for random-pattern resistant faults using some deterministic algorithm and storing them in a Read-Only Memory (ROM)~\cite{SaDB84}. The latter approach can help detecting not only random-pattern resistant faults, but also delay faults which are not handled efficiently by the pseudo-random patterns. However, the memory required to store the top-off patterns in BIST can exceed 30\% of the memory used in a conventional ATPG approach~\cite{HeF99}. Finding alternative ways of generating top-off patterns is an important open problem.

\begin{figure}[t] %f|_{x_i=0} \not = f|_{x_i=1}\},
	\centering
	\begin{subfigure}[t]{\columnwidth}
		\centering
		\includegraphics[width=2.5in]{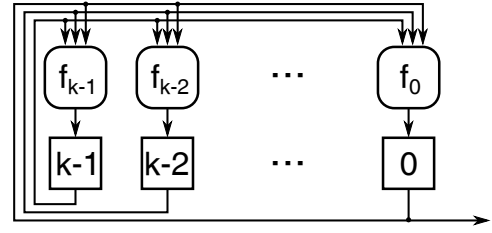}
		\caption{An RNLU with the degree of parallelization one.}
		\label{fig:bm}
	\end{subfigure}
	\vbox to 10pt{}
	\begin{subfigure}[t]{\columnwidth}
		\centering
		\includegraphics[width=2.5in]{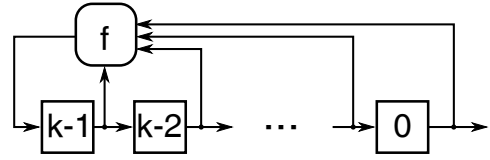}
		\caption{An NLFSR with the degree of parallelization one.}
		\label{fig:fsr}
	\end{subfigure}
	\caption{General structure of RNLUs and NLFSRs.}
	\label{fig:fsr_bm}
\end{figure}

Any binary sequence can be generated using a Register with Non-Linear Update (RNLU) shown in Figure~\ref{fig:bm}.
A $k$-stage RNLU consists of $k$ binary stages, $k$ updating functions, and a clock. At each clock cycle, the current values of all stages are synchronously updated to the next values computed by the updating functions.  
RNLUs can be viewed as a more general type of 
Non-Linear Feedback Shift Registers (NLFSRs) (see Figure~\ref{fig:fsr})
in which both, feedback and feedforward, connections are allowed and no chain connection between the stages is required.

RNLUs are typically smaller and
faster than NLFSRs generating the same sequence. 
For example, consider the 4-stage NLFSR with the updating function 
\[
f(x_0,x_1,x_2,x_3) = x_0 \oplus x_3 \oplus x_1 \cdot x_2 \oplus x_2 \cdot x_3,
\]
where ``$\oplus$'' is the Boolean exclusive-OR, ``$\cdot$'' is the Boolean AND,
and $x_i$ is the variable representing the value of the stage $i$, 
$i \in \{0,1,2,3\}$.
If this NLFSR is initialized to the state $(x_3 x_2 x_1 x_0) = (0001)$, it 
generates the output sequence 
\begin{equation} \label{bs}
(1,0,0,0,1,1,0,1,0,1,1,1,1,0,0)
\end{equation}
with the period 15.
The same sequence can be generated by the 4-stage RNLU with the 
updating functions
\[
\begin{array}{lcl}
f_3(x_0,x_3) & = & x_0 \oplus x_3 \\
f_2(x_1,x_2,x_3) & = & x_3 \oplus x_1 \cdot x_2  \\
f_1(x_2) & = & x_2 \\
f_0(x_1) & = & x_1.
\end{array}
\]
We can see that the RNLU uses 3 binary operations,
while the NLFSR uses 5 binary operations. 
%Furthermore, the depth of feedback functions of the RNLU is smaller than
%the depth of the feedback function of the NLFSR. 
%Thus, the RNLU is faster than the NLFSR.

While RNLUs can potentially be smaller than NLFSRs, 
the search space for finding a smallest RNLU for a given sequence
is considerably larger than the corresponding one for NLFSRs.
Algorithms for constructing RNLUs  with the minimum number of stages
were presented in~\cite{Du10aj,Du11a}. 
However, since, for large $k$, the size of a circuit implementing a $k$-input Boolean function is typically much larger than the size of a single stage of a register, usually these algorithms do not 
minimize the total size of an RNLU.

In this paper, we present an algorithm which minimizes the size of the support set of updating functions, i.e. the number of variables on which the updating functions depend.
For most Boolean functions, the size of a circuit computing a function grows exponentially with the number of the variables in their support set~\cite{Sh49}. Therefore, by
reducing the number of variables of updating functions to the minimum, we
can minimize the total size of an RNLU. To support this claim,
we derive expressions for the expected size of RNLUs constructed by the presented method and previous approaches. Our analysis shows that RNLUs constructed by the presented method are asymptotically smaller.
For completeness, we also compare RNLUs to linear and nonlinear feedback shift registers generating the same sequence.

The rest of this paper is organized as follows. 
Section~\ref{sec:prel} lists the notation and basic concepts used in the paper.
Section~\ref{sec:prev} discusses the related work.
Section~\ref{sec:idea} gives a general introduction to the presented approach.
Section~\ref{sec:alg} describes the algorithm for constructing RNLU.
Section~\ref{sec:ana} compares RNLUs constructed by the presented method 
to the RNLUs constructed using previous approaches, as well as to linear and nonlinear feedback shift registers.
Section~\ref{sec:exp} presents the experimental results.
Section~\ref{sec:con} concludes the paper.

\section{Preliminaries}
\label{sec:prel}

In this section, we present basic definitions and notation used in the paper.

\subsection{Boolean functions}

A $k$-variable \emph{Boolean function} is a mapping of type $f: B^k \to B$, where $B = \{0,1\}$.
The \emph{support set} of a Boolean function $f(x_0,x_1,\cdots,x_{k-1})$, $sup(f)$, is a set of variables on which $f$ depends:
\[
sup(f) = \{x_i \ | \ f|_{x_i=0} \not = f|_{x_i=1}\},
\]
where $f|_{x_i=j} = f(x_0, \cdots, x_{i-1}, j, x_{i+1}, \cdots, x_{k-1})$, for $j \in \{0,1\}$.
 
A $k$-variable Boolean function $f$ can be computed by a {\em logic circuit} with $k$ inputs and one output, such that, for every input combination $a \in B^k$, the circuit output is $f(a)$.
%An \emph{implementation} of $m$ $k$-variable Boolean functions $f_0,f_1,\cdots,f_{m-1}$, is a combinational circuit with $k$ inputs and $m$ outputs, that for every input combination $\mathbf{x}$, the circuit outputs $f_0(\mathbf{x}),f_1(\mathbf{x}),\cdots,f_{m-1}(\mathbf{x})$.
The \emph{size} of a circuit is the number of gates required to implement it.
%~\cite{Bo77}.
Typically gates are restricted to a certain set, e.g. $\{$AND, OR, NOT$\}$~\cite{We87}.
%It circuit-size complexity gives us the theoretical lower bound 
%on the size of the implementation of a Boolean function, using the same set of gates.

\subsection{Registers with Non-Linear Update}

%The name "RNLU" was introduced by Golomb in his seminal book~\cite{Golomb_book}.

A $k$-stage Register with Non-Linear Update (RNLU) (also called {\em binary machine}~\cite{Golomb_book,Du10aj}) consists of $k$
binary storage elements, called {\em stages}, each capable of storing one bit of information. 
Every stage $i \in \{0,1, \cdots,n-1\}$ has an associated \emph{state variable} $x_i \in \{0,1\}$ 
which represents the current value of the stage $i$ and a Boolean \emph{updating function} 
$f_i: \{0,1\}^k \rightarrow \{0,1\}$ which determines how the value of $x_i$ is updated
to its next value, $x_i^+$:
\[
x_i^+ = f_i(x_0,x_1,\cdots,x_{k-1}).
\]

A {\em state} of an RNLU is a vector of values of its
state variables. At every clock cycle, 
the next state of an RNLU is computed from its the current state 
by updating the values of all stages simultaneously
to the values of the corresponding updating functions.

%The \emph{size} of a $k$-stage RNLU is equal to the sum of sizes of implementations of its $k$ updating functions and the size of its $k$ storage elements. 
%In our analysis, we take a 2-input Boolean gate for a unit of size. We assume that all 2-input Boolean gates have the same size and 
%a single-bit storage element is larger than a 2-input gate by a constant.

%One or more of stages of an RNLU may be used for producing the output sequence. 
The \emph{degree of parallelization} $p$ of a $k$-stage RNLU is the number of  
stages used for producing the output at each clock cycle, $1 \leq p \leq k$.
Throughout the paper, we assume that 
$p$ rightmost stages of RNLU are used for producing its output.

\subsection{Feedback Shift Registers}

A $k$-stage \emph{Feedback Shift Register} (FSR) can be viewed as a special case
of a $k$-stage RNLU satisfying
$$
\begin{array} {rcl}
x_0^+ &=& x_1 \\
x_1^+ &=& x_2 \\
&\cdots& \\
x_{k-2}^+ &=& x_{k-1} \\
x_{k-1}^+ &=& f(x_0,x_1,\cdots,x_{k-1})
\end{array} 
$$
The updating function of the stage $k-1$ is called the \emph{feedback function} of the FSR.
%The structure of an FSR is shown in Figure~\ref{fig:fsr}.

If all feedback functions of an FSR are linear, then the FSR is called a {\em Linear Feedback Shift Register} (LFSR).
Otherwise, it is called a {\em Non-Linear Feedback Shift Register} (NLFSR).

Its is known that the recurrence relation 
generated by the feedback function of a $k$-stage LFSR  
has a characteristic polynomial of degree $k$~\cite{Golomb_book}. If this polynomial is primitive
\footnote{An
irreducible polynomial of degree $k$ is called {\em primitive} if 
the smallest $m$ for which it divides $x^m + 1$ is equal to $2^k - 1$~\cite{LiH94}.}, then the LFSR follows a periodic sequence of $2^k-1$ states 
which consists of all possible non-zero $k$-bit vectors~\cite{Golomb_book}.
This result is very important, because it makes possible the generation of pseudo-random sequences of length $2^k-1$ with a 
device of size $O(k)$. No analogous results has been found for the
nonlinear case yet.

\section{Previous Work} 
\label{sec:prev}

There are many different ways of generating binary sequences. A thorough treatment of this topic is
given by Knuth in~\cite{Kn69}. In this section, we focus on FSR-based binary sequence generators and their generalizations.

LFSRs are one of the most popular devices for generating 
pseudo-random binary sequences. They have numerous applications, including
error-detection and correction~\cite{McCluskey1999},  data compression~\cite{MrRT04}, 
testing~\cite{Da98}, and cryptography~\cite{MuS06}.

The Berlekamp-Massey algorithm can be used to construct
a smallest LFSR generating a given binary sequence. It was originally invented by
Berlekamp for decoding Bose-Chaudhuri-Hocquenghem (BCH) codes~\cite{Be67}. 
Massey~\cite{Ma69} linked the Berlekamp's algorithm to LFSR synthesis and simplified it.
There were many subsequent extensions and improvements of the algorithm,
for example Mandelbaum~\cite{Ma84} developed its arithmetic analog, 
Imamura and Yoshida~\cite{ImY87}  presented an alternate and easier derivation,
Fitzpatrick~\cite{Fi94}  found a version which is  
more symmetrical in its treatment of the iterated pairs of polynomials,
and Fleischmann~\cite{Fl95} modified it to extend the model sequence in both directions around any given data bit. 
It has also been shown that similar to the Berlekamp-Massey algorithm results can be obtained with 
the Euclidean algorithm~\cite{Do87} and continued fractions~\cite{WeS79}.

The Berlekamp-Massey algorithm constructs traditional LFSRs, which generate one output bit per clock cycle.
A number of techniques have been developed for constructing LFSRs with the degree of parallelization $p$.
Two main approaches are: (1) synthesis of subsequences representing $p$ decimation of some phase shift of the original LFSR sequence~\cite{Lempel1971} and (2) computation of the set of states reachable from any state in $p$ steps. The latter is usually done by computing 
$p$th power of the connection matrix of the LFSR~\cite{MuS06}.
LFSRs with a high degree of parallelization are used in applications where high data rate is important, such a Cyclic Redundancy Check (CRC)  widely used in data transmission 
and storage for detecting burst errors~\cite{McCluskey1999}.

NLFSRs have been much less studied compared to LFSRs~\cite{Fr82}.
The first algorithm for constructing a smallest NLFSR generating a given binary sequence
was presented by Jansen in 1991~\cite{Ja89,Ja91}. Alternative algorithms were 
given by  Linardatos et al~\cite{603256}, Rizomiliotis et al~\cite{RiK05}, and Limniotis et al~\cite{LiKK07}.

Similarly to the LFSR case, an NLFSR can be re-designed to 
generate $p$ bits of the sequence per clock cycle.
This is usually done by duplicating the updating functions of an NLFSR $p$ times, as in~\cite{canniere-trivium,hell-grain,cryptoeprint:2005:415}.
Such a technique requires that the $p$ left-most stages of the NLFSR are not used as inputs to feedback functions or output functions.
More generally, the problem of constructing an NLFSR with the degree of parallelization $p$ can be solved
by computing the $p$th power of the transition relation induced by its feedback functions.
However, the size of circuits computing the $p$th power of the transition relation may grow substantially larger than a factor of $p$~\cite{DuS12}.

An FSR may need up to $n$ stages to generate a binary sequence of length $n$.
For example, the smallest LFSR and NLFSR generating the binary sequence 
\[
\underbrace{0 0 \cdots 0}_{n-1} 1,
\]
have $n$ and $n-1$ stages, respectively~\cite{Ja89}.

On average, an LFSR needs $n/2$ stages to generate a binary sequence of length $n$~\cite{Ru86}
and an NLFSR needs $2\log_2{n}$ stages to generate such a sequence~\cite{Ja89}. 
Note that these bounds reflect the size of stages only; they do not take into account the size of circuits computing feedback functions. Since nonlinear feedback function of an NLFSR is typically larger than the linear feedback function of an LFSR, a $k$-stage NLFSR may be considerably larger than a $k$-stage LFSR. 

The first algorithm for constructing an RNLU with the minimum number of stages for a given binary sequence was presented in~\cite{Du10aj}. 
This algorithm exploits the unique property of RNLUs that \emph{any} binary $n$-tuple can be the next state of a given current state. 
The algorithm assigns every 0 of a sequence a unique even integer and every 1 of a sequence a unique odd integer.
Integers are assigned in an increasing order starting from 0. 
For example, if an 8-bit sequence $A= (0,0,1,0,1,1,0,1)$ is given, the sequence of integers (0,2,1,4,3,5,6,7) can be used. 
This sequence of integers is interpreted as a sequence of states of an RNLU. 
The largest integer in the sequence of states determines the number of stages. 
In the example above, $\lceil \log_2 7 \rceil = 3$, thus the resulting RNLU has 3 stages. 
 
In~\cite{Du11a}, the algorithm~\cite{Du10aj} was extended to RNLUs generating $p$ bits 
of the output sequence per clock cycle. 
The main idea is to encode a binary sequence into an $2^p$-ary sequence which can be generated by a smaller RNLU. 
As an example, suppose that we use the 4-ary encoding $(00) = 0, (01) = 1, (10) = 2, (11) = 3$ to encode the binary sequence $A$ from the example above, into the  quaternary sequence (0,2,3,1).
Then, we can construct an RNLU generating the sequence $A$ 2-bits per clock cycle using a sequence of states (0, 2, 3, 1). 
Note that $\lceil \log_2 3 \rceil = 2$, so the resulting RNLU has one stage less than the RNLU generating one bit per clock cycle in the previous example. 

%RNLUs are very efficient in generating parallel deterministic sequences.
%It is shown in~\cite{Du11a} that RNLUs outperform the previous methods in generating sequences of bit vectors.

RNLUs have been successfully applied to the storage of cryptographic keys~\cite{LiSD13} and deterministic test patterns~\cite{LiD13tr}. For example, it was shown in~\cite{LiSD13} that an RNLU may take less than a quarter of the size of a read-only memory storing the same sequence.

%, RNLUs have been use for the storage of secret keys.
%The efficiency of the algorithm is improved through introducing don't-care values into state encoding.
%The experimental results show that RNLUs consume less chip area than ROMs when storing sequences with fewer than 128K bits.

\section{Intuitive Idea}
\label{sec:idea}

\begin{figure}[t]
	\centering
	\includegraphics[width=1.4in]{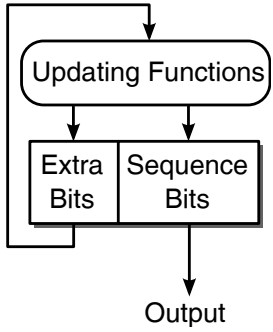}
	\caption{Structure of RNLUs constructed by the presented algorithm.}
	\label{fig:rbm}
\end{figure}

We can separate each state of a $k$-stage RNLU with the degree of parallelization $p$ into two parts: $p$ \emph{output bits} which contain the output sequence and
$k-p$ \emph{extra bits} which are used for differentiating the states whose output bits are the same.
Output bits are defined by the sequence to be generated. 
For the extra bits, we can use any $k-p$ bit vector that is not used in another state with the same output bits.

As we mentioned previously, the overall size of an RNLU is typically dominated by the size of circuits computing its updating functions. The size of these circuits greatly depends on the support sets of updating functions. 
In order to minimize the support sets, we use extra bit vectors which are unique for every specified state.
In other words, not only the states with the same output bits, but also all other specified states are assigned a unique $(k-p)$-bit extra bit vector.
Such a state encoding allows us to reduce the support sets of updating functions to variables representing extra bits only, as shown in Figure~\ref{fig:rbm}.

Suppose we would like to construct an RNLU generating a binary sequence $A$ of length $m \times p$ with the degree of parallelization $p$. In order to distinguish between identical $p$-bit vectors in $A$,
we need at least $\lceil \log_2 m \rceil$ extra bits.
Therefore, the number of stages in the resulting RNLU is given by:
\[
k = \lceil \log_2 m \rceil + p.
\]
This number is typically greater than the minimum possible number of stages in an RNLU which can generate $A$. The minimum number of stages is determined by partitioning $A$ into $p$-bit vectors,
computing the decimal representation for each $p$-bit vector, and counting the largest number of occurrences among all $p$-bit vectors with the same decimal representation, $N_{max}$. For example, in the 10-bit sequence $A = (0,1,0,0,0,1,1,1,0,1)$ the 2-bit vector (0,1) occurs 3 times, so $N_{max} = 3$. The minimum number of stages in an RNLU generating $A$ is given by~\cite{Du11a}:
\begin{equation} \label{kmin}
k_{min} = \lceil \log_2 N_{max} \rceil + p.
\end{equation}

The presented method reduces the support sets of the updating functions to the minimum. 
Updating functions of output bits cannot depend on less than $\lceil \log_2 m \rceil$ 
variables since otherwise the RNLU would not be able to generate all
$\lceil n/p \rceil$ $p$-bit vectors constituting a partitioning of $A$.
%the RNLU would not be able to generate an (aperiodic) sequence 
%of $m$ consecutive output vectors.
%\footnote{Throughout the paper, we assume that the sequence to be generated is aperiodic.}. 

Note that the size of an RNLU can be further reduced by removing the stages representing output bits and taking the output directly from the updating functions. 

%However, if the output is directly fed into another combinational block, the critical path delay may increase, limiting the maximum clock frequency of the entire circuit.

\section{Algorithm}
\label{sec:alg}

In this section, we present an algorithm for constructing RNLUs which minimizes the support sets 
of updating functions to $\lceil \log_2 m \rceil$ variables representing extra bits. 

The pseudocode of the algorithm \texttt{ConstructRNLU}($A,p$) is shown as Algorithm~\ref{alg1}. The input is a binary sequence $A = (a_0, a_1, \cdots, a_{n-1})$
and the desired degree of parallelization $p$.
The output is the defining tables of $p+r$ updating functions 
of the RNLU generating $A$ with the degree of parallelization $p$,
where $r = \lceil \log_2{m} \rceil$ and $m = \lceil n/p \rceil$.

The algorithm begins by selecting an $r$-stage extra bits generator $G$
using the procedure \texttt{ChooseGenerator}($n,r$).
As we mentioned in the previous section, the size of an RNLU depends on the order 
of extra bit vectors used for state encoding.
In principle, any permutation of $r$-bit vectors can be used, however, a good choice
of the generator reduces the size of the resulting RNLU.
For example, if we use an $r$-stage LFSR or a binary counter as generators of
extra bit vectors, then the updating functions of extra bits can be computed by a circuit of size 
$O(r)$. 

The selected generator $G$ is set to some initial state $g_0 \in B^r$. For LFSRs, $g_0$ must be a non-zero state. For binary counters, $g_0$ can be any state.
Then, the defining table of updating functions of output bits is constructed as follows. 
At every step $i$, $i \in \{0,1,\cdots,m-1\}$,
the input part of the table is assigned to be the current state of the generator $G$, $g_i$,
and the output part of the table is assigned to be the $i$th $p$-bit vector of the input sequence $A$.

All remaining $2^r-m$ input assignments are mapped to don't-care values.
This gives us a possibility to specify the functions $f_0, f_1, \cdots, f_{p-1}$
so that the size of their circuits is minimized.
%Finally, the set of resulting updating functions is returned.

Since, by construction, the values of functions $f_0, f_1, \cdots, f_{p-1}$
at step $i$
correspond to the $i$th
$p$-tuple of $A$, for $i \in \{0,1,\cdots,m-1\}$, the resulting RNLU generates $A$ 
with the degree of parallelization $p$.

\begin{algorithm}[t!]
\caption{\texttt{ConstructRNLU}($A,p$)
Constructs an RNLU generating a binary sequence $A = (a_0, a_1, \cdots, a_{n-1})$ 
with the degree of parallelization $p$.}
\label{alg1}
\begin{algorithmic}[1]
\STATE $m = \lceil n/p \rceil$;
\STATE $r = \lceil \log_2{m} \rceil$;
\STATE $G =$ \texttt{ChooseGenerator}($m,r$); 
\STATE Initialize $G$ to an initial state $g_0 \in B^r$;
    \FOR{every $i$ from 0 to $m-1$}
        \FOR{every $j$ from 0 to $p-1$}
            \STATE $f_j(g_i) = a_{i*p+j}$;
        \ENDFOR
        \STATE $g_{i+1} =$ \texttt{ComputeNextState}($G,g_i$);
    \ENDFOR

    \FOR{every $i$ from 0 to $r-1$}
	\STATE $f_{p+i} =$ updating function of the stage $i$ of $G$;
    \ENDFOR
\STATE Return  $f_0, f_1, \cdots, f_{p+r-1}$;
\end{algorithmic}
\end{algorithm}

As an example, let us construct an RNLU 
which generates the following 40-bit binary sequence with the degree of parallelization 4:

\[
\begin{array}{r}
A = (1,0,0,1, 0,0,1,0, 0,0,1,1, 0,0,1,0, 1,0,1,0, 1,0,1,0, \\
0,0,0,1, 1,0,0,0, 0,1,1,0, 1,1,1,0)
\end{array}
\]

%First, we partition $A$ into 10 4-bit vectors:
%\[
%(1001, 0010, 0011, 0010, 1010, 1010, 0001, 1000, 0110, 1110)
%\]
We need $r = \lceil \log_2 10 \rceil = 4$ extra bits to assign to each 
of the 10 4-bit vector of $A$ a unique extra bit vector.
Suppose that we use the 4-stage LFSR 
with the primitive generator polynomial $g(x) = 1 + x + x^4$ for generating
extra bits.
If we choose (0001) as the initial state of the LFSR, then  
extra bit vectors are assigned according to 
the following sequence of LFSR states:
\[
(1, 8, 4, 2, 9, 12, 6, 11, 5, 10).
\]
This gives us the following defining table for the updating
functions of output bits:
\[
\begin{tabular}{|c|cccc|} \hline
$x_7 x_6 x_5 x_4$ & $f_3$ & $f_2$ & $f_1$ & $f_0$  \\ \hline
0~0~0~1 & 1 & 0 & 0 & 1 \\
1~0~0~0 & 0 & 1 & 0 & 0 \\
0~1~0~0 & 1 & 1 & 0 & 0 \\
0~0~1~0 & 0 & 1 & 0 & 0 \\
1~0~0~1 & 0 & 1 & 0 & 1 \\
1~1~0~0 & 0 & 1 & 0 & 1 \\
0~1~1~0 & 1 & 0 & 0 & 0 \\
1~0~1~1 & 0 & 0 & 0 & 1 \\
0~1~0~1 & 0 & 1 & 1 & 0 \\
1~0~1~0 & 0 & 1 & 1 & 1 \\ \hline
\end{tabular}
\]
%where ``-'' stands for a don't care value.
These functions can be implemented as follows:
\begin{eqnarray*}
f_3(x_7,x_6,x_5,x_4) &=& \overline{x}_7(\overline{x}_5 + x_6)(\overline{x}_4 + x_5 + \overline{x}_6) \\
f_2(x_7,x_6,x_5,x_4) &=& (x_7 + (x_5 \oplus x_6))(\overline{x}_4 + \overline{x}_5 + x_6 + \overline{x}_7) \\
f_1(x_7,x_6,x_5,x_4) &=& (\overline{x}_4 + \overline{x}_7)(x_6 + x_7)(x_5 \overline{x}_6 + x_4 \overline{x}_5) \\
f_0(x_7,x_6,x_5,x_4) &=& x_4 x_7 + (x_7 \oplus \overline{x}_5 \overline{x}_6)
\end{eqnarray*}
where ``$+$'' is the Boolean OR and $\overline{x}$ denotes the Boolean complement of $x$.

The updating functions of extra bits, $f_7, f_6, f_5, f_4$ are defined by the LFSR: 
\[
\begin{array}{lcl}
f_7(x_4,x_5) & = & x_4 \oplus x_5\\
f_6(x_7) & = & x_7\\
f_5(x_6) & = & x_6\\
f_4(x_5) & = & x_5
\end{array}
\]

\begin{algorithm}[t!]
\caption{\texttt{ChooseGenerator}($m,r$)
Chooses an $r$-stage generator of extra bits with at least $m$ states.}
\label{alg2}
\begin{algorithmic}[1]
\IF{$m < 2^r$}
    \STATE $G =$ Any $r$-stage LFSR with a primitive generator polynomial of degree $r$;
\ELSE
    \STATE $G =$ $r$-stage binary counter;
\ENDIF   
\STATE Return  $G$;
\end{algorithmic}
\end{algorithm}

\begin{figure}[t]
	\centering
    \includegraphics[width=2.9in]{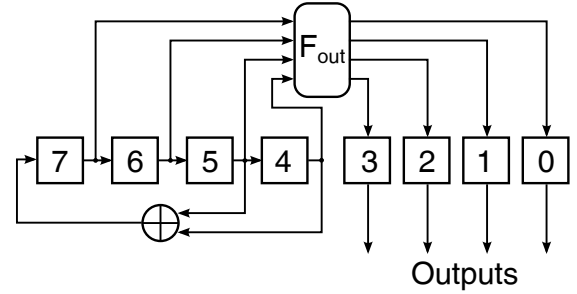}
	\caption{8-stage RNLU constructed for the example.}
	\label{fig:circ_ex}
\end{figure}

Figure~\ref{fig:circ_ex} shows the structure of the resulting RNLU.
The block labeled by $F_{out}$ computes the updating functions of output bits $f_3, f_2, f_1, f_0$.

%\subsection{Treatment for Incompletely Specified Sequences}
%
%In the presented method, incompletely specified sequences are handled gracefully.
%Because sequence bits only appear in the output part of the truth table of $F_\mathrm{out}$, the don't-care bits in the sequence can remain unspecified in the truth table.
%In this case, every bit in the sequence takes a value from the set $\{0,1,d\}$, instead of $\{0,1\}$, and the rest of the algorithm remains the same.

%%%% TODO: finish the example
%As an example, consider

\section{Expected Size Analysis}
\label{sec:ana}

In this section, we derive expressions for the expected size of RNLUs constructed using the presented algorithm and the algorithms~\cite{Du10aj} and~\cite{Du11a}.
For completeness, we also show results for LFSRs and NLFSRs generating the same sequence.

In 1942, Shannon~\cite{Sh49} has proved that there is an (asymptotically) large fraction of Boolean functions
of $k$ variables that remains uncomputable with circuits of size larger than $2^k/k$.  
In 1962, Lupanov~\cite{Lu62} has shown that, if we allow circuit size to be larger by a small fraction of $2^k/k$, namely $[1 + o(1)]  2^k/k$, then we can compute all $k$-variable Boolean functions. In both cases, it is assumed that circuits are composed from AND, OR and NOT gates with at most two inputs.

From these two bounds, we can conclude that ``most'' Boolean function of $k$ variables require a 
circuit of size $\alpha 2^k/k$ to be computed, where $\alpha$ is a constant such that $1 \leq \alpha \leq 2$.

In the analysis below, we assume one storage element counts as $\beta$ gates.
Since the analysis is asymptotic, without the loss of precision we use $\log_2{n}$ instead of $\lceil \log_2{n} \rceil$.  

\subsection{Degree of Parallelization One}

Let $A$ be a binary sequence of length $n$ in which every element is selected independently and uniformly at random from $B$. Throughout this section, we call such a sequence a {\em random sequence}. 
Suppose that Algorithm 1 is used to construct an RNLU generating $A$ with the degree of parallelization one. Then, the resulting RNLU has:
\begin{itemize}
\item one stage for the output bit,
\item $\log_2{n}$ stages for extra bits,
\item $\log_2{n}$ updating functions of the extra bits,
\item one updating function of the output bit.
\end{itemize}

The updating functions of the extra bits can be computed by a circuit of size $O(\log_2{n})$. 
%The updating function of the output bit has the support set of size $\log_2{n}$, because the probability that there exists an extra bit variable $x_i$ such that $x_i \not \in sup(f_0)$ goes to 0 as the sequence length increases.
The updating function $f_0$ of the output bit is expected to depend on all $\log_2{n}$ state variables of extra bits. This is because the probability that 
$f_0|_{x_i=0} = f_0|_{x_i=1}$ for some $i \in \{1,2,\cdots,(\log_2{n})-1\}$
goes to 0 as the sequences length increases.
Therefore, $f_0$ requires a circuit of size
$\alpha n / \log_2{n}$ to be computed. So, the expected size of the RNLU constructed by the presented algorithm is

\begin{eqnarray}
E[RNLU(n,1)] &=& \beta (1 + \log_2{n}) + \alpha n / \log_2{n} + O(\log_2{n}) \nonumber\\
           &=& O(n / \log_2{n}). \label{eq:rbm}
\end{eqnarray}

Next, suppose that the algorithm~\cite{Du10aj} is used to construct an RNLU for the same sequence.
This algorithm constructs an RNLU with the minimum number of stages $k_{min}$ given by~(\ref{kmin}). For sufficiently large random sequences, this number can be approximated as:
\[
k_{min} \approx 1 + \log_2(n/2) = \log_2{n}s.
\]

In this case, the resulting RNLU has $k_{min}$ stages and $k_{min}$ updating functions with the support set of size $k_{min}$. These functions required $k_{min}$ circuits of size $\alpha 2^{k_{min}} / k_{min}$ to be computed, so their expected size is given by:
\[
k_{min} \cdot \alpha 2^{k_{min}} / k_{min} = \alpha 2^{\log_2{n}} = \alpha n.
\]
Therefore, the expected size of the RNLU constructed by the algorithm~\cite{Du10aj} is:

\begin{eqnarray}
E[RNLU(n,1)] = \alpha n + \beta \log_2{n} = O(n). \label{eq:bm} 
\end{eqnarray}

Next, suppose that Berlekamp-Massey algorithm~\cite{Ma69} is used to construct an LFSR for the same sequence. Suppose that this LFSR has $l$ stages. According to~\cite{Ru86}, for sufficiently large random sequences, $l \approx n/2$. The linear feedback function of the LFSR can be computed by a circuit of size $O(n)$. So, the expected size of the LFSR is
\begin{eqnarray}
E[LFSR(n,1)] = \beta n / 2 + O(n) = O(n). \label{eq:lfsr}
\end{eqnarray}

Finally, suppose an $r$-stage NLFSR is constructed of the same sequence, e.g. using the algorithm~\cite{603256}.
According to~\cite{Ja89}, for sufficiently large random sequences, $r \approx 2\log_2{n}$.
Thus, the feedback function of the NLFSR has the support set of size $2\log_2{n}$.
It requires a circuit of size $\alpha \cdot 2^{2\log_2{n}}/(2\log_2{n})$ to be computed.
Therefore, the expected size of the NLFSR is
\begin{eqnarray}
E[NLFSR(n,1)] \nonumber &=& 2 \beta \log_2{n} + \alpha \cdot 2^{2\log_2{n}}/(2\log_2{n}) \nonumber\\
&=& 2 \beta \log_2{n} + \alpha n^2 / (2\log_2{n}) \nonumber\\
&=& O(n^2 / \log_2{n}). \label{eq:nlfsr}
\end{eqnarray}

As we can see from equations (\ref{eq:rbm}), (\ref{eq:bm}), (\ref{eq:lfsr}), and (\ref{eq:nlfsr}), for sufficiently large random sequences, RNLUs with the degree of parallelization one constructed by the presented algorithm are asymptotically smaller than RNLUs constructed by the algorithm~\cite{Du10aj}, LFSRs, and NLFSRs.

\subsection{Degree of Parallelization $p$}

In this section, we extend the analysis to the degree of parallelization $p$.

Let $A$ be a random binary sequence of length $n$. Suppose that Algorithm 1 is used to construct an RNLU generating $A$ with the degree of parallelization $p$. Let $m = \lceil n/p \rceil$. Then this RNLU has:
\begin{itemize}
\item $p$ stages for the output bits,
\item $\log_2{m}$ stages for extra bits,
\item $\log_2{m}$ updating functions of the extra bits,
\item $p$ updating functions of the output bits.
\end{itemize}

The updating functions of the extra bits can be computed by a circuit of size $O(\log_2{m})$. 
%Each of the $p$ updating functions of the output bits has the support set of size $\log_2{m}$, because for any $j \in \{0,1,\cdots,p-1\}$, the probability that there exists an extra bit variable $x_i$ such that $x_i \not \in sup(f_j)$ goes to 0 as the sequence length increases.
Each of the $p$ updating functions of the output bits is expected to depend on all $\log_2{m}$ state variables of extra bits. This is because, for any $j \in \{0,1,\cdots,p-1\}$, the probability that $f_j|_{x_i=0} = f_j|_{x_i=1}$ for some $i \in \{p,p+1,\cdots,(p+\log_2{m})-1\}$
goes to 0 as the sequences length increases.
Therefore, the updating functions of output bits require $p$ circuits of size
$\alpha m / \log_2{m}$ to be computed. Thus, the expected size of the RNLU constructed by the presented algorithm is
\begin{eqnarray}
E[RNLU(n,p)] &=& \beta (p + \log_2{m}) + p \alpha m / \log_2{m} + O(\log_2{m}) \nonumber\\
&=& O(n / \log_2{m}) \nonumber\\
&=& O(n / \log_2{(n/p)}). \label{eq:rbmp}
\end{eqnarray}

Suppose that the algorithm~\cite{Du11a} is used to construct an RNLU for the same sequence.
The number of stages $k_{min}$ is given by~(\ref{kmin}).
Since $1 \leq N_{max} \leq m$, we get 
\[
p \leq k_{min} \leq p + \log_2{m}.
\]
The lower bound is reached when each $p$-bit vector occurs in $A$ exactly once. This
is possible only if $n \leq 2^p$. Therefore
\begin{equation} \label{km}
\log_2{n} \leq k_{min} \leq p + \log_2{m}.
\end{equation}

The $k_{min}$ updating functions require $k_{min}$ circuits of size $\alpha 2^{k_{min}} / k_{min}$ to be computed, so their expected size is $\alpha 2^{k_{min}}$.
From~(\ref{km}), we get:
\[
\alpha n \leq  \alpha 2^{k_{min}} \leq \alpha m 2^p
\]
Therefore, the lower bound on expected size of the RNLU constructed by the algorithm~\cite{Du11a} is:
\begin{eqnarray}
E[RNLU(n,p)] & \geq & \beta \log_2{n} + \alpha n  \nonumber\\
           & \geq & O(n).\label{eq:bmp}
\end{eqnarray}

An LFSR with the degree of parallelization $p$ has the same number of stages as the LFSR with the degree of parallelization one, but its feedback function is modified to compute $p$th power of the connection matrix. This implies that the expected size of the circuit computing the feedback function of the LFSR  increases $p$ times. So, the expected size of the LFSR is
\begin{eqnarray}
E[LFSR(n,p)] = \beta n / 2 + O(pn) = O(pn). \label{eq:lfsr_p}
\end{eqnarray}

Similarly, NLFSRs with the degree of parallelization $p$ are constructed by modifying its feedback functions to compute $p$th power of its transition relation. This may increase in the size of the circuit computing $p$th power of its transition relation more than $p$ times due to multiplication of non-linear terms~\cite{DuS12}. The the expected size of the NLFSR is thus
\begin{eqnarray}
E[NLFSR(n,p)] \nonumber & \geq & 2 \beta \log_2{n} + \alpha \cdot p \cdot 2^{2\log_2{n}}/(2\log_2{n}) \nonumber\\
&\geq & 2 \beta \log_2{n} + \alpha  p  n^2 / (2\log_2{n}) \nonumber\\
&\geq & O(p  n^2 / \log_2{n}). \label{eq:nlfsr_p}
\end{eqnarray}

From equations (\ref{eq:rbmp}), (\ref{eq:bmp}), (\ref{eq:lfsr_p}), and (\ref{eq:nlfsr_p}), we can conclude that, for sufficiently large random sequences, RNLUs with the degree of parallelization $p$ constructed by the presented algorithm are asymptotically smaller than RNLUs constructed by the algorithm~\cite{Du10aj}, LFSRs, and NLFSRs. 

Note that our analysis does not take into account that two circuits implementing two $k$-variable Boolean functions may share some gates, and therefore their cost may be smaller than $2 \alpha 2^k/k$.
However, since the analysis is asymptotic, this factor is not likely to affect the results. 

\section{Experimental Results}
\label{sec:exp}

\begin{figure}[t]
	\centering
	\begin{subfigure}[t]{\columnwidth}
		\centering
		\includegraphics[width=2.5in]{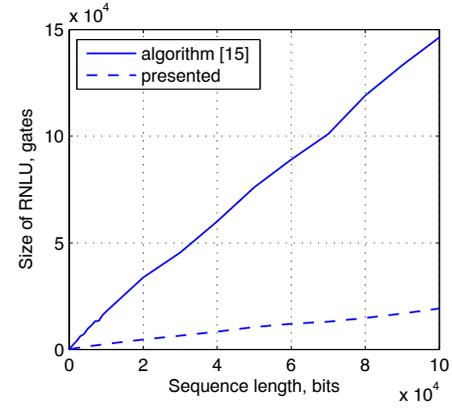}
		\caption{Degree of parallelization one.}
		\label{fig:exp-w1}
	\end{subfigure}
	\begin{subfigure}[t]{\columnwidth}
		\centering
		\includegraphics[width=2.5in]{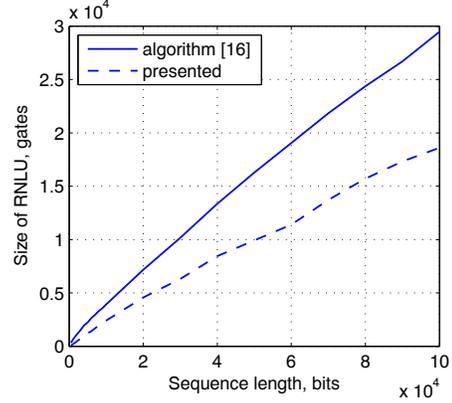}
		\caption{Degree of parallelization 100.}
		\label{fig:exp-w100}
	\end{subfigure}
	\caption{Comparison of RNLUs constructed by the presented algorithm to RNLUs constructed using the algorithms~\cite{Du10aj,Du11a}. Each dot is computed as an average for 100 randomly generated sequences of the same length.}
	\label{fig:exp}
\end{figure}

To compare the analytical results to the actual size of RNLUs, we
applied the presented algorithm and algorithms~\cite{Du10aj, Du11a}, 
to randomly generated binary sequences of length up to $10^5$ bits.

%Binary counters were used to generate extra bits for the presented method.
For all algorithms, circuits for the updating functions were synthesized 
using the logic synthesis tool ABC~\cite{abc}. 
The generic library of gates \emph{mcnc.genlib} was used for technology mapping. 

Figures~\ref{fig:exp-w1} and \ref{fig:exp-w100} show the results for the 
degrees of parallelization 1 and 100, respectively. 2-input AND is used as 
a unit of gate size.
We can see that RNLUs constructed by the presented algorithm are considerably smaller that RNLUs
constructed by the algorithms~\cite{Du10aj} and~\cite{Du11a}.
The improvement is particularly striking for the degree of parallelization one.
For example, for sequences of length $10^5$, 
RNLUs constructed by the algorithm~\cite{Du10aj} are 6.67 times larger
than RNLUs constructed by the presented algorithm.
For the degree of parallelization 100 and sequences of length $10^5$, 
RNLUs constructed by the algorithm~\cite{Du11a} are 65.1\% larger
than RNLUs constructed by the presented algorithm.

\section{Conclusion}
\label{sec:con}

In this paper, we presented an algorithm for constructing RNLUs 
in which the support set of updating functions
is reduced to the minimum. We proved that the expected size of the resulting RNLUs is 
asymptotically smaller than the expected size of RNLUs constructed by previous approaches.
%We have also shown that our results are within $O(\log_2(n/p))$ gates from optimal.

The presented method might be useful for applications which require efficient generation of
binary sequences, such as testing, wireless communication, and cryptography.

%Future work includes a more thorough evaluation of reduced RNLU, and applying the technique in potential applications.

%\appendix[Estimating the Number of Stages in an RNLU]
%
%Let an RNLU generating sequence $S$ of width $p$ and length $n$ be synthesized by algorithm~\cite{Du11a}.
%Then, the number of stages is given by
%\[
%k = p + \lceil \log2{N_\mathrm{max}} \rceil,
%\]
%where $N_\mathrm{max}$ is the maximum number of occurences of any word in the sequence~\cite{Du11a}.
%
%Assuming each word in the sequence is purely random and independant, the probability for a word to be a specific value is $2^{-p}$.

\section*{Acknowledgement}
This work was supported in part by the research grant No 2011-03336 
from Swedish Governmental Agency for Innovation Systems
(VINNOVA) and in part by the research grant No 621-2010-4388 from the Swedish 
Research Council.

\balance
\bibliographystyle{ieeetr}
\bibliography{bib}  

\end{document}